\documentclass[12pt]{article}

\usepackage{preprintstyle}
\usepackage[utf8]{inputenc}

\newcommand{\beq}{\begin{equation}}
\newcommand{\eeq}{\end{equation}}
\newcommand{\bal}{\begin{aligned}}
\newcommand{\eal}{\end{aligned}}
\newcommand{\rmd}{\mathrm d}

\usepackage{physics, xparse, tikz}
\usetikzlibrary{decorations.pathmorphing, decorations.pathreplacing, decorations.shapes}
\usepackage{subfig}
\usetikzlibrary{shapes.geometric}
\usepackage{dsfont}
\usepackage[normalem]{ulem}
\usepackage{cancel,xcolor}

\title{Modular Fluctuations in Cosmology}

\author{Lars Aalsma$^{a,b}$,}
\author{Sang-Eon Bak$^{b}$}
\emailAdd{laalsma@asu.edu}
\emailAdd{sbak2@asu.edu}
\affiliation{$^a$Beyond: Center for Fundamental Concepts in Science, Arizona State University, Tempe, Arizona 85287, USA}
\affiliation{$^b$Department of Physics, Arizona State University, Tempe, Arizona 85287, USA}

\abstract{Verlinde and Zurek (VZ) have proposed that quantum gravity fluctuations in causal diamonds lead to observable effects. In particular, they argued that in quantum gravity causal diamonds have an uncertainty in their size that scales as $\delta L \sim \sqrt{\ell_p L}$, i.e. fluctuations are enhanced from the Planck scale. In this work, we explore the origin of the VZ scaling by studying scalar perturbations in inflationary cosmology. We reinterpret these cosmological perturbations as fluctuations of the modular Hamiltonian associated with a causal diamond, establishing a connection between modular fluctuations and observable primordial perturbations. Independent of assumptions about the modular Hamiltonian or the background spacetime, our results show that---whenever an (effective) spherically symmetric perturbation is quantized---it can lead to the VZ scaling. However, whether such a mode exists in the vacuum of any theory of quantum gravity remains an open question.}

\begin{document}

\maketitle

\section{Introduction}
Quantum gravity effects are typically associated with Planck scale physics. Since the Planck length is extremely small, i.e. $\ell_p \sim 10^{-35}\, \rm m$, effects of this size are likely unobservable. At the same time, there is ample evidence that challenges the notion that quantum gravity effects are always confined to the Planck scale. One example where quantum gravity has macroscopic consequences is the black hole information paradox: deviations from semi-classical gravity propagate to distance scales that are much larger than the Planck scale and involve the scale of the horizon. A recent work along these lines is  \cite{Bousso:2023kdj}.

However, it has been debated whether these large-distance quantum gravity effects also have implications on `simple' observables. By simple, we mean observables that are directly probed by a realistic experiment. Clearly, the black hole information paradox and in particular confirming unitarity of black hole radiation does not fall into this category. This would requires carefully measuring the quantum state of about $N\propto S_{\rm BH} $ Hawking quanta, where $S_{\rm BH}$ is the black hole entropy: an endeavor that does not seem feasible with current technology.

This leads us to search for other opportunities where quantum gravity might leave macroscopic signatures. A few years ago, this search was revitalized\footnote{Previous ideas along these lines have been explored, e.g. \cite{Hogan:2007pk}, but experimental efforts have only resulted in constraints \cite{Holometer:2017ovp}.} in work by Verlinde and Zurek \cite{Verlinde:2019xfb}. In that paper, it was proposed that the modular Hamiltonian, an operator associated to the density matrix of a spacetime subregion, can exhibit fluctuations that backreact on the geometry. While such fluctuations are not surprising by itself, \cite{Verlinde:2019xfb} claimed that this results in an uncertainty in the size of a causal diamond that scales as $\delta L^2 \sim \ell_p L$. If so, quantum gravity fluctuations in a causal diamond are enhanced from the Planck scale by the size of the causal diamond. We will refer to this scaling as the Verlinde-Zurek (VZ) effect. Further works studying this effect include \cite{Verlinde:2019ade,Zurek:2020ukz,Banks:2021jwj,Gukov:2022oed,Verlinde:2022hhs,Li:2022mvy,Zhang:2023mkf,Bub:2023bfi,He:2023qha,Lee:2023kry,He:2024ddb,Bak:2024kzk,He:2024skc,He:2024vlp,Bub:2024nan,Banks:2024cqo}.

Despite this large body of work, it is fair to say the claimed scaling has remained controversial. In particular, a recent work \cite{Carney:2024wnp} studied the roundtrip time of photons reflected at a mirror in the presence of a quantized gravitational wave. This models the single arm of an interferometer and \cite{Carney:2024wnp} found that the arm length fluctuation scales as $\delta L^2 \sim \ell_p^2$, reaffirming conventional wisdom and suggesting that the VZ effect must stem from more exotic quantum gravity effects. However, the fundamental difference between these different approaches has remained unclear up to now.

The goal of this paper is to take a step towards answering this question. We will do so by studying cosmological inflation. As we will explain in detail, we can think of primordial perturbations in inflationary cosmology as quantum fluctuations of a causal diamond in quasi-de Sitter space.\footnote{Quasi-de Sitter space refers to an Friedmann-Robertson-Walker (FRW) geometry that is approximately de Sitter space in the sense that we can perturb the scale factor around exact de Sitter space.} The novelty of our approach lies in the fact that we show how to reinterpret these (known) perturbations as modular Hamiltonian fluctuations, by proposing an ansatz for the modular Hamiltonian of a quasi-de Sitter background. Because primordial scalar perturbations during inflation are directly related to temperature fluctuations in the Cosmic Microwave Background (CMB), the upshot of this approach is a direct link between modular Hamiltonian fluctuations and observations.

Importantly, irrespective of assumptions about the modular Hamiltonian we show that in out setup correlators of proper length and time are consistent with the scaling proposed by Verlinde and Zurek. Moreover, we argue that this scaling is generic, and in particular also holds for causal diamonds flat space. We show that this scaling stems from quantizing a spherically symmetric mode of the gravitational field. We contrast this with the approach taken by \cite{Carney:2024wnp} where a radiative mode of the gravitational field was quantized. This explains the different scaling of correlators of proper time obtained in \cite{Carney:2024wnp}.

Our work therefore clarifies how the VZ scaling can come about: by quantizing an (effective) spherically symmetric perturbation in quantum gravity. Because our cosmological setup involves a matter contribution, the s-wave of the inflaton plays this role in inflation. It is unclear to us, however, whether such a mode is also present in the (quantum gravity) vacuum.\footnote{In the classical theory, such a mode can be gauged away. This does not preclude its existence in quantum gravity however.} If so, our results support the idea that quantum gravity fluctuations might be observable at interferometers.

The rest of this paper is organized as follows. In Sec. \ref{sec:Cosmology} we review some basic facts about inflationary cosmology and show how primordial scalar perturbations can be related to modular Hamiltonian fluctuations. Independent of assumptions about the modular Hamiltonian, we study the impact of fluctuations on correlators of proper distance and time in Sec. \ref{sec:Diamonds}. We discuss our results in Sec. \ref{sec:Discussion} and summarize details about the coordinate systems we used in Appendix \ref{app:Coordinates} and relate different gauges we use to study perturbations in Appendix \ref{app:Gauge}.

\section{Modular Hamiltonian Fluctuations during Inflation} \label{sec:Cosmology}
In this section, we first review classical inflationary backgrounds and include linear perturbations in the metric as well as the stress tensor. We then quantize those perturbations. Using an ansatz for the modular Hamiltonian we show how fluctuations in the modular Hamiltonian can be related to quantized metric perturbations. To set up our inflationary background, we found the references \cite{Maggiore:2018sht,Baumann:2022mni} to be especially helpful.

\subsection{Classical Friedmann-Robertson-Walker Cosmology}
The background geometry of the early universe is to a good approximation homogeneous and isotropic and described by the Friedmann-Robertson-Walker line element.
\beq \label{eq:PlanarUnperturbed}
\rmd s^2 = a(\eta)^2\left(-\rmd \eta^2 + \rmd \vec x^2\right) = a(\eta)^2\left(-\rmd \eta^2 + \rmd \rho^2 + \rho^2\rmd\Omega_2^2\right) ~.
\eeq
where we are working with the conformal time $\eta\in(-\infty,0)$, and $\rmd\Omega_2^2$ describes a round two-sphere. Because of the flat spatial slices, we will refer to this coordinate system as planar coordinates. In planar coordinates, due to the conformal factor the geometry is expanding as the proper distance $\zeta$ between two events with coordinate (or comoving) distance $\rho = \sqrt{\delta_{ij}x^ix^j}$ scales with time as $\zeta=a(\eta)\rho$.

In the very early universe, during inflation, the background is approximately described by de Sitter space. This means that the scale factor is approximately equal to its value in exact de Sitter space. Qualitatively, this means that we take $a(\eta) = -(H(\eta)\eta)^{-1}$ and assume the Hubble parameter $H(\eta)$ to only depend weakly on time, as $H$ is a constant in de Sitter space. We also find it useful to define the so-called conformal Hubble parameter ${\cal H}(\eta) = a'(\eta)/a(\eta) = H(\eta)a(\eta)$, where a $'$ denotes a derivative with respect to $\eta$. Following standard terminology we refer to this inflationary background as quasi-de Sitter space.

The time dependence of the Hubble parameter can be made concrete by a so-called slow-roll expansion. In terms of the coordinate $\rmd\tau= a(\eta)\rmd\eta$ we define the slow-roll parameter as
\beq \label{eq:epsilon_def}
\varepsilon = -\frac{\partial_\tau H}{H^2}  = 1 - \frac{{\cal H}'}{{\cal H}^2} ~.
\eeq
In the exact de Sitter limit $\varepsilon \to 0$. Quasi-de Sitter space corresponds to a geometry where $|\varepsilon|\ll 1$. In this article, we always work to the leading non-trivial order in $\varepsilon$.

In inflation, the expansion in quasi-de Sitter space is sourced by a homogeneous and isotropic scalar field $\bar\varphi(\eta)$ with potential energy $V(\bar \varphi)$. The energy density $\bar{\rho} = - T^\eta_{\,\,\,\eta}$ and pressure $\bar{P} = \frac13 T^i_{\,\,\,i}$ of this scalar field are given by
\beq
\bal
\bar{\rho} & = +V(\bar \varphi) + \frac12\left(\frac{\bar\varphi'(\eta)}{a(\eta)}\right)^2 ~, \\
\bar{P} &= -V(\bar \varphi) + \frac12\left(\frac{\bar\varphi'(\eta)}{a(\eta)}\right)^2 ~.
\eal
\eeq
To drive accelerated expansion, the potential energy needs to dominate over kinetic energy, i.e. $\bar{\rho} =  - \bar{P} + {\cal O}(\varepsilon)$. To leading order in $\varepsilon$ this is the equation of state of a cosmological constant.

Because the background geometry only slowly evolves with time it is appropriate to think of each stage of inflation as being described by a de Sitter universe with a slightly different Hubble parameter $H$. Doing so, we can introduce static coordinates in which the metric \eqref{eq:PlanarUnperturbed} is given by\footnote{See Appendix \ref{app:Coordinates} for an overview of the different coordinate systems we use and their relations.}
\beq
\rmd s^2 = b(t,r)^2\left(-\left(1-H^2r^2\right)\rmd t^2 + \left(1-H^2r^2\right)^{-1}\rmd r^2 + r^2\rmd\Omega_2^2\right) ~.
\eeq
Here, we defined $b^2 = (H\eta a)^2$. For quasi-de Sitter space $b^2 =1 + {\cal O}(\varepsilon)$.

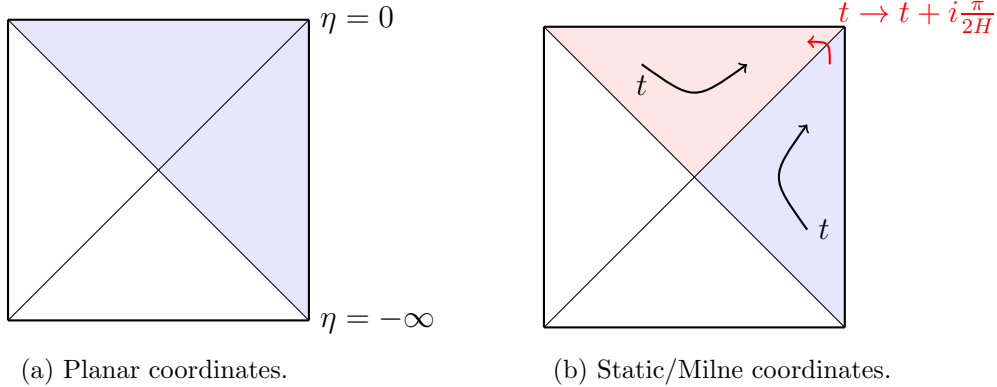
\begin{figure}[t]
\centering
\subfloat[\qquad\qquad\qquad\qquad]{{
\begin{tikzpicture}

\draw [draw=white] (0,-.2) rectangle (6,5);

\node[isosceles triangle,
    isosceles triangle apex angle=90,
    fill=blue!10,
    rotate=45,
    minimum size =2.8cm] at (2.8,2.8){};

\draw[thick] (0,0) -- (4,0);
\draw[thick] (0,0) -- (0,4);
\draw[thick] (0,4) -- (4,4);
\draw[thick] (4,0) -- (4,4) node[pos=0,right]{$\eta=-\infty$} node[pos=1,right]{$\eta=0$};

\draw (0,0) -- (4,4);
\draw (0,4) -- (4,0);

\end{tikzpicture}
}}
    \qquad
    \subfloat[\qquad\qquad\qquad]{{

\begin{tikzpicture}

\draw [draw=white] (0,-.2) rectangle (6,5);

\node[isosceles triangle,
    isosceles triangle apex angle=90,
    fill=blue!10,
    rotate=180,
    minimum size =2cm] (T90)at (3.2,2){};

\node[isosceles triangle,
    isosceles triangle apex angle=90,
    fill=red!10,
    rotate=-90,
    minimum size =2cm] (T90)at (2,3.2){};

\draw[thick] (0,0) -- (4,0);
\draw[thick] (0,0) -- (0,4);
\draw[thick] (0,4) -- (4,4);
\draw[thick] (4,0) -- (4,4);
\draw (0,0) -- (4,4);
\draw (0,4) -- (4,0);

\draw[->,thick] (3.5,1.3) .. controls (3,2) .. (3.5,2.7) node[pos=0,right]{$t$};
\draw[->,thick] (1.3,3.5) .. controls (2,3) .. (2.7,3.5) node[pos=0,below]{$t$};

\draw[->,thick, red] (3.8,3.5) .. controls (3.8,3.8) .. (3.5,3.8) node[pos=0.5,above right]{$t\to t+i\frac{\pi}{2H}$};

\end{tikzpicture}

    }}
    \caption{Penrose diagrams of de Sitter space with the shaded regions covered by (a) planar coordinates and (b) static coordinates (in blue) and Milne coordinates (in red).}
    \label{fig:PenroseDiagrams}
\end{figure}

Static coordinates only cover a quarter of the Penrose diagram of global de Sitter space. We can analytically continue the time coordinate to describe different quarters of the Penrose diagram. In particular, we will make use of this fact to also describe the Milne wedge (the topmost quarter), see Figure \ref{fig:PenroseDiagrams}.
In the static metric it is clear that there is a horizon located at $r_h=H^{-1}$. In planar coordinates, the horizon corresponds to $\eta+\rho=0$.

Having specified the background geometry we now include linear perturbations. On the metric side, it is useful to decompose the different possible perturbations by using their transformation properties under spatial rotations. This scalar-vector-tensor decomposition is convenient in cosmology because it separates the equations of motion for each of the perturbations.

For our purposes, we will only consider scalar perturbations and set the other perturbations to zero. The most general perturbed metric is then given by (see e.g. \cite{Baumann:2022mni})
\beq
\rmd s^2 = a^2(\eta)\left(-(1+2A)\rmd\eta^2 + 2B_i\rmd x^i\rmd\eta+(\delta_{ij}+2E_{ij})\rmd x^i\rmd x^j\right) ~,
\eeq
with
\beq
\bal
B_i &= \partial_iB ~, \\
E_{ij} &= C\delta_{ij} + \left(\partial_i\partial_j-\delta_{ij}\square_{(3)}\right)E  ~,
\eal
\eeq
where $A,\,B,\,C$, and $E$ depend on space and conformal time. Here $\square_{(3)}$ is the Laplacian on flat Euclidean three-space. The metric contains four scalar degrees of freedom $(A,B,C,E)$, but only two of them are physical. Indeed, writing the metric as $g_{\mu\nu} = \bar g_{\mu\nu} + h_{\mu\nu}$, with $\bar g_{\mu\nu}$ the background metric, we can perform a linear diffeomorphism of the form
\beq 
h_{\mu\nu} \to h_{\mu\nu} - (\bar\nabla_\mu\xi_\nu+\bar\nabla_\nu\xi_\mu) ~.
\eeq
The two (physical) scalar degrees of freedom that are invariant under this diffeomorphism are the Bardeen potentials \cite{Bardeen:1980kt}
\beq
\bal
\Psi &= A + {\cal H}(B-E') + (B-E')' ~, \\
\Phi &=-C + \frac13\square_{(3)}E-{\cal H}(B-E') ~.
\eal
\eeq
We can now use the gauge freedom to remove the two gauge degrees of freedom. We find it convenient to work in conformal Newtonian gauge, which amounts to setting $B=E=0$. In this gauge, the metric takes the form
\beq \label{eq:PlanarPerturbed}
\rmd s^2 = a(\eta)^2\left(-\left(1+2\Psi\right)\rmd\eta^2 + \left(1-2\Phi\right)\left(\rmd\rho^2 + \rho^2\rmd\Omega_2^2\right)\right) ~,
\eeq
where we used spherical coordinates. As before, we can transform to static coordinates to find
\beq
\bal \label{eq:static-perturbed}
\rmd s^2 &= b(t,r)^2\left(-f\rmd t^2 + g^{-1}\rmd r^2 + R^2\rmd\Omega_2^2+ \frac{4Hr(\Psi+\Phi)}{1-H^2r^2}\rmd t\rmd r\right)
\eal
\eeq
with
\beq \label{eq:ShiftedParametersPlanar}
\bal
f &= \left(1-H^2r^2(1-2\Phi)+2\Psi\right) ~, \\
g &= \left(1-H^2r^2(1-2\Psi)+2\Phi\right) ~,\\
R^2 &= r^2(1-2\Phi) ~. \\
\eal
\eeq
In the absence of anisotropic stress, which is the case for inflation, the equations of motion set $\Psi = \Phi$. Using this, we find that the remaining perturbation $\Phi$ has several effects on the static line element. First, to linear order in $\Phi$ it shifts the root of $f(r)$ indicating a shift in the horizon.\footnote{We give the explicit location of the horizon in an FRW geometry in Sec. \ref{subsec:Thermo}.} Second, it introduces an off-diagonal term that, in static coordinates \cite{Christensen:1977jc}, indicates a flux of energy through the horizon. In fact, it can be shown that this flux obeys thermodynamic identities that are equivalent to the usual Friedmann equations governing the evolution of the inflationary solution \cite{Frolov:2002va}.

For the perturbed geometry to solve the Einstein equations, we also need to perturb the stress tensor. Writing the scalar field perturbations as $\varphi(\eta,\vec x) = \bar\varphi(\eta) + \delta \varphi(\eta,\vec x)$ the perturbed stress tensor can be found by taking the usual definition
\beq
T_{\mu\nu} = \partial_\mu\varphi\partial_\nu\varphi - g_{\mu\nu}\left(\frac12(\partial \varphi)^2+V(\varphi)\right) ~,
\eeq
and expanding it to linear order in $\delta\varphi$.

We will now discuss the Einstein equations in planar coordinates. The time-time component and off-diagonal component of the Einstein equations for the perturbations now yield
\beq
\bal
3{\cal H}^2\Phi+3{\cal H}\Phi' - \nabla_{\vec x}^2\Phi &= 4\pi G_N\left({\bar\varphi{}'}^2\Phi-\bar\varphi'\delta\varphi'-a^2(\partial_{\bar\varphi}V(\bar\varphi))\delta\varphi\right) ~, \\
\Phi'+{\cal H}\Phi &= 4\pi G_N\bar\varphi'\delta\varphi ~.
\eal
\eeq
Here $\nabla_{\vec x}^2 = g^{ij}\nabla_i\nabla_j$. In addition, we will make use of the background Einstein equations to obtain the useful expression
\beq \label{eq:phi_prime_express}
\bar\varphi'^2 = \frac{{\cal H}^2}{4\pi G_N}\varepsilon ~,
\eeq
where we used the definition \eqref{eq:epsilon_def}.

By making a judicious change of variables, these equations can be manipulated to yield the Mukhanov-Sasaki equation: a `master' equation that determines the evolution of the scalar perturbation we are studying. A useful (gauge-invariant) variable is the comoving curvature perturbation ${\cal R}$, given by \cite{Baumann:2009ds}
\beq \label{eq:Rexpress}
{\cal R} = \Phi + {\cal H}\left(\frac{\delta\varphi}{\bar\varphi'}\right) ~.
\eeq
Using the perturbed Einstein equations and \eqref{eq:phi_prime_express}, we can derive an equation for the derivative of ${\cal R}$.
\beq \label{eq:Rderiv}
{\cal R}' = \frac{\cal H}{4\pi G_N\bar \varphi'^2}\nabla_{\vec x}^2\Phi ~.
\eeq
By transforming to momentum space, this equation tells us that ${\cal R}$ is constant in the long-wavelength limit. 

Next, we introduce the so-called Mukhanov-Sasaki variable $u=z{\cal R}$ with $z =a \bar\varphi'/{\cal H}$. Using the equation of motion for the background scalar field, given by
\beq
\varphi'' +2{\cal H}\varphi' + a^2 \partial_{\bar\varphi}V(\bar\varphi) = 0 ~,
\eeq
we can eliminate the derivative of the potential from the time-time component of the Einstein equations. Also using the off-diagonal component we can then derive
\beq \label{eq:PhiExpress}
\Phi' + {\cal H}\Phi = \frac{4\pi G_N}{a} z{\cal H}\delta\varphi ~.
\eeq
Using \eqref{eq:Rexpress} and the change of variables to $u$, we can obtain
\beq
\Phi = z^{-1}\left(u-a\delta\varphi\right) ~.
\eeq
Plugging this into \eqref{eq:PhiExpress}, we now only have to eliminate $\delta\varphi$ and its derivatives. This can be achieved by using \eqref{eq:Rderiv} and rewriting it in terms of $u$. Doing so finally leads to the Mukhanov-Sasaki equation:
\beq
u'' - \left(\frac{z''}{z}+\nabla_{\vec x}^2\right)u = 0 ~.
\eeq
We note that this equation is valid for any FRW geometry, not just quasi-de Sitter space.

To solve this equation, it will be useful to transform to momentum space. We therefore write the modes just in terms of the magnitude of the momentum $|\vec k| = k$ and the equation for the momentum space variable $u_k$ is given by
\beq \label{eq:MSequation}
u_k''+\left(k^2 - \frac{z''}{z}\right)u_k = 0 ~.
\eeq
If we focus on modes with long wavelengths, i.e. $k^2 \ll |z''/z|$ the general solution to \eqref{eq:MSequation} is
\beq
u_k(z) = u_+ z + u_- z\int\frac{\rmd \eta}{z^2} ~, 
\eeq
where $u_+$ and $u_-$ are integration constants. We are mostly interested in the solution proportional to $u_+$, because it corresponds to a constant comoving curvature perturbation ${\cal R}_k= u_+$ if we define the momentum space variable ${\cal R}_k = u_k/z$ for the comoving curvature perturbation.

Let's focus on quasi-de Sitter space. By using the definition of $z$ and \eqref{eq:phi_prime_express}, we find that
\beq
z^2 = \frac{a^2}{4\pi G_N}\varepsilon ~,
\eeq
We therefore see that the solution proportional to $u_-$ decays with time and only the constant solution remains. In the exact de Sitter limit $\varepsilon\to 0$, $z''/z = 2/\eta^2$ and we see that ${\cal R}_k$ `freezes out' (becomes constant) at superhorizon scales $k|\eta|\ll 1$. This fact leads to the observability of temperature anisotropies in the CMB.

Similar manipulations can be performed directly on $\Phi$ to show that the classical solution $\Phi_k$ also becomes constant on superhorizon scales in quasi-de Sitter space, see e.g. \cite{Mukhanov:1985rz}. Next, we will quantize these perturbations.

\subsection{Quantization of Scalar Perturbations} \label{sec:quantization}
We now quantize the Mukhanov-Sasaki variable $u$ and relate it to the metric perturbation $\Phi$. Promoting $u$ to an operator, we expand it as
\beq \label{eq:mode-expansion}
\hat u(\eta,\vec x) = \int\frac{\rmd ^3k}{(2\pi)^3}\left(u_k(\eta)\hat a_{\vec k}+u_k^*(\eta)\hat a^\dagger_{-\vec k}\right)e^{i\vec k\cdot \vec x} ~.
\eeq
The $u_k(\eta)$ is a solution to the Mukhanov-Sasaki equation and, as usual, we choose the boundary conditions that correspond to the Bunch-Davies vacuum annihilated by $\hat a_k$.
\beq \label{eq:MSsolution}
u_k(\eta) = \frac{e^{-ik\eta}}{\sqrt{2k}}\left(1-\frac{i}{k\eta}\right) ~.
\eeq
Instead of the plane wave basis used in \eqref{eq:mode-expansion} we can also expand in spherical harmonics. This is useful because we will be focusing on spherically symmetric perturbations. We therefore make the replacement
\beq
e^{i\vec k\cdot\vec x} = 4\pi \sum_{l=0}^\infty\sum_{m=-\ell}^{\ell} i^\ell j_{\ell}(k\rho)Y_{\ell m}(\hat k)Y_{\ell m}^*(\hat \rho) ~.
\eeq
Here $j_{\ell}(k\rho)$ is the spherical Bessel function, $\rho^2 = \delta_{ij}x^ix^j$ and $Y_{\ell m}(\hat k)$ are spherical harmonics. In the long-wavelength limit $j_{\ell}(k\rho) \sim (k\rho)^{\ell}$ so we see that all modes with $\ell>0$ decay and only the s-wave survives. 

We note that the solution \eqref{eq:MSsolution} is valid for arbitrary $k$. By remembering that $u_k=z{\cal R}_k$ has a straightforward relation with the comoving curvature perturbation and by employing the identity \eqref{eq:Rderiv} in momentum space:
\beq\label{eq_RPhirel}
{\cal R}_k' = \frac{k^2}{{\cal H}\varepsilon}\Phi_k ~,
\eeq
correlation functions of $\hat u(\eta,\rho)$ can now be easily related to correlation functions of $\hat {\cal R}(\eta,\rho)$ or $\hat \Phi(\eta,\rho)$.

Concretely, let us consider the zero-point fluctuations. To be precise, we will hereafter take the state to be the Bunch-Davies vacuum state. Using the mode expansion \eqref{eq:mode-expansion}, we find that
\beq
\langle{\cal R}(\eta,\rho)^2\rangle = \int \frac{\rmd ^3k}{(2\pi)^3}|{\cal R}_k(\eta)|^2 = \int_0^\infty\rmd k\frac{k^2}{2\pi^2}|{\cal R}_k(\eta)|^2
\eeq
Defining the dimensionless power spectrum $\Delta^2_{\cal R}$ as
\beq
\langle{\cal R}^2(\eta,\rho)\rangle = \int_0^\infty\frac{\rmd k}{k}\Delta^2_{\cal R}(\eta) ~,
\eeq
and using the solution \eqref{eq:MSsolution} and $u_k= z\mathcal{R}_k$ we obtain in the long-wavelength limit
\beq
\Delta^2_{\cal R} = \frac{G_NH^2}{2\pi\varepsilon} ~.
\eeq
This is the well-known result for the inflationary power spectrum, which is constant in the limit $(k\eta)^2\to 0$.

We can now compute the zero-point fluctuations of $\hat \Phi(\eta,\rho)$, just focusing on the s-wave sector. Using the relation between $\Phi_k$ an ${\cal R}_k$ \eqref{eq_RPhirel}, we obtain
\beq
\Delta^2_{\Phi} = \frac{G_NH^2\varepsilon}{2\pi} = \frac{\varepsilon}{2S_{\rm dS}} ~,
\eeq
where we defined the de Sitter entropy as the area of the horizon divided by $4G_N$.
\beq
S_{\rm dS} = \frac{\text{Area}}{4G_N} = \frac{\pi}{G_NH^2} ~.
\eeq
Notably, the power spectrum $\Delta^2_{\Phi}$ is independent of $k$, constant and vanishes in the exact de Sitter limit $\varepsilon\to 0$. The last property is a reflection of the fact that in exact de Sitter space, these perturbations become pure gauge modes associated with the exact time-translation symmetry of the background, see e.g. \cite{Jarnhus:2007ia}.

Returning to the variance itself, we find that
\beq \label{eq:DivergentInt}
\langle {\cal R}(\eta,\rho)^2 \rangle = \frac{1}{2\varepsilon S_{\rm dS}}\int_0^\infty\frac{\rmd k}{k} ~,
\eeq
is divergent both in the UV and IR. The UV divergence is the usual divergence associated to (flat space) short distance physics whereas the IR divergence (and similar ones in general de Sitter correlation functions) has been subject to debate. Recent advances, see \cite{Green:2022ovz} for a recent review of (some of these) issues, have demonstrated how to deal with these IR divergences. Here, we follow the approach of \cite{Huenupi:2024ksc} and use dimensional regularization. This leads to (see Eq. (43) of \cite{Huenupi:2024ksc})
\beq
\langle {\cal R}^2(\eta,\rho) \rangle = \frac{1}{2\varepsilon S_{\rm dS}}\log(\mu/H) ~,
\eeq
where $\mu$ is a renormalization scale. The integral appearing in the variance of $\Phi$ contains the same logarithmic divergence and can similarly be renormalized to yield
\beq \label{eq:RenormalizedVariance}
\langle \Phi^2(\eta,\rho) \rangle = \frac{\varepsilon}{2S_{\rm dS}}\log(\mu/H) ~,
\eeq
In the next subsection, we will see how quantum fluctuations in $\Phi$ can be related to fluctuations in the modular Hamiltonian.

\subsection{Thermodynamics and the Modular Hamiltonian} \label{subsec:Thermo}
We will now relate fluctuations of the modular Hamiltonian to the primordial scalar fluctuations we studied in the previous subsection. To do so, we will need to define the modular Hamiltonian associated with the causal region accessible to an observer in the expanding spacetime.

If we define a quantum field theory on a fixed de Sitter background, the reduced density matrix of the static patch, the largest causal diamond, takes a thermal form in the Bunch-Davies state:
\beq
\sigma_{\rm dS} = \frac1{Z}\exp(-\beta \hat H) ~.
\eeq
Here $Z$ is a normalization factor, $\beta = 2\pi/H$ is the inverse de Sitter temperature and $\hat H$ is the Hamiltonian generating time translation with the static time coordinate $t$. In this simple case, the (dimensionless) modular Hamiltonian, up to the normalization, is given by
\beq
K_{\rm dS} = -\log(\sigma_{\rm dS}) = \beta\hat H ~,
\eeq
and coincides with the static patch Hamiltonian. From the definition of the modular Hamiltonian we see that the modular Hamiltonian obeys
\beq
\langle K_{\rm dS}\rangle = -\text{tr}\left(\sigma_{\rm dS}\log\sigma_{\rm dS}\right) = S ~,
\eeq
where $S$ is the entanglement entropy of the density matrix $\sigma_{\rm dS}$. In field theory, the entanglement entropy is divergent. 

Now consider what happens when we turn on dynamical gravity. In this case, recent works on the von Neumann (vN) algebra of observables in de Sitter space \cite{Chandrasekaran:2022cip,Frob:2023hwf} have revealed that in gravitational systems the type III vN algebra associated to quantum field theory is modified to a type II vN algebra.\footnote{To obtain a non-trivial algebra in de Sitter space requires introducing an observer to break the time translation invariance, but in inflationary geometries the inflaton itself already does this \cite{Chen:2024rpx}. Related recent work \cite{Geng:2024dbl} further explores the role of spacetime translation symmetry breaking in the construction of gauge-invariant operators.} The upshot is that this allows one to rigorously show that the entanglement entropy of some subregion equals the (finite) generalized entropy:
\beq
S_{\rm gen} = \frac{\text{Area}(\partial \Sigma)}{4G_N} + S_{\rm vN}(\Sigma) ~,
\eeq
where $S_{\rm vN}(\Sigma)$ is the von Neumann entropy of matter fields in the subregion $\Sigma$, whose domain of dependence is the static patch.\footnote{To be precise, the divergent piece of the von Neumann entropy can be absorbed by the renormalized Newton's constant \cite{Susskind:1994sm}.} For the static patch of de Sitter space we therefore find that the expectation value of the modular Hamiltonian is given by
\beq \label{eq:ModHam_dS}
\langle K_{\rm dS} \rangle = \frac{\pi}{G_NH^2} + S_{\rm vN} ~.
\eeq
A similar result holds for inflationary backgrounds, which have been studied in \cite{Kudler-Flam:2024psh,Chen:2024rpx}.

Before we discuss the relation between the modular Hamiltonian and scalar fluctuations we studied before, let us consider various thermodynamic quantities in a general FRW background. Because a general FRW background does not have an exact timelike Killing vector, defining the horizon and its associated thermodynamics is more involved than in pure de Sitter space. 

However, working in planar coordinates we consider spheres of constant $R=a(\eta)\rho$ and define the expansion of those spheres. The character of the normal vector $n_\mu = \partial_\mu R$ determines whether the spherical three-volume bounded by $R$ is untrapped, marginal or trapped \cite{Hayward:1997jp}. This allows us to define the horizon as the surface beyond which the expansion becomes negative.

Given the unperturbed planar metric in polar coordinates the normal vector to a surface of constant $R$ is
\beq
n = a^{-1}\left(-\rho{\cal H}\partial_\eta +\partial_\rho\right) ~.
\eeq
Since we are interested in the location of a horizon, we will look for the surface where the norm of $n^\mu$ vanishes, which is given by
\beq
n^\mu n_\mu = 0 \quad \to \quad \rho_h = 1/{\cal H} ~.
\eeq
This is not a Killing horizon, but due to the spherical symmetry we can define a more general Kodama vector $k^\mu$ \cite{Kodama:1979vn} that becomes null at $\rho_h$. In spherically symmetric spacetimes, the Kodama vector can be used to define a current that is locally conserved \cite{Kinoshita:2024wyr}. It is defined to be orthogonal to the normal vector, i.e. $k^\mu n_\mu = 0$ and its norm satisfies
\beq
k^\mu k_\mu = \frac{2G_NM(R)}{R} - 1 ~,
\eeq
where $M(R)$ is the energy contained in the volume bounded by $R$. This quantity is known as the Misner-Sharp mass \cite{Misner:1964}. Thus, using the Kodama vector we can define a generalized notion of mass for FRW spacetimes. Explicitly,
\beq
k = a^{-1}\left(\partial_\eta + \rho{\cal H}\partial_\rho\right) ~.
\eeq
Evaluated at the horizon, the Misner-Sharp mass is given by
\beq
M(R_h) = \frac{R_h}{2G}= \frac{1}{2GH} ~.
\eeq
In addition, we also define a generalized notion of surface gravity $\kappa$ which obeys \cite{Hayward:1997jp}
\beq
\frac12k^\mu\left(\nabla_\nu k_\mu - \nabla_\mu k_\nu\right) = \kappa n_\nu ~,
\eeq
From this equation we find on the horizon that
\beq
\kappa = H\left(1-\frac{\varepsilon}2\right) ~.
\eeq
This suggests an identification of the temperature of the horizon by $T=\kappa/(2\pi)$ which reduces to the de Sitter temperature in the $\varepsilon\to 0$ limit. We also note that in this limit the Misner-Sharp mass obeys
\beq
M(R_h) = TS_{\rm dS} ~,
\eeq
with $S_{\rm dS}$ the de Sitter entropy. Comparing this with \eqref{eq:ModHam_dS}, this relation shows that in (quasi-)de Sitter space the Misner-Sharp mass can be identified with the leading ${\cal O}(G_N^{-1})$ term in the expectation value of the modular Hamiltonian, up to a normalization with the temperature.
\beq
\langle K\rangle = \beta M(R_h) + {\cal O}(G_N^0) ~.
\eeq
Now, we will add perturbations. Including perturbations, the planar line element in polar coordinates is
\beq
\rmd s^2 = a(\eta)^2\left(-\left(1+2\Phi\right)\rmd\eta^2 + \left(1-2\Phi\right)\left(\rmd\rho^2+\rho^2\rmd\Omega_2^2\right)\right) ~.
\eeq
Using the same logic as before, the normal vector to a surface of constant $R^2=(a(\eta)\rho)^2(1-2\Phi)$ and Kodama vector can now be used to define the Misner-Sharp mass and surface gravity. The vectors are
\beq
\bal
n &= a^{-1}\left({\cal H}\rho\left(-1 + 3\Phi\right) + \rho\Phi'\right)\partial_\eta +a^{-1}\left(1+\Phi-\rho\partial_\rho\Phi\right)\partial_\rho ~,\\
k &= a^{-1}\left(1-\Phi-\rho\partial_\rho\Phi\right)\partial_\eta-a^{-1}\left(\rho{\cal H}(1-\Phi)-\rho\Phi'\right)\partial_\rho ~.
\eal
\eeq
The shifted location of the horizon is then given by
\beq \label{eq:ShiftedHor}
n^\mu n_\mu = 0 \quad \to \quad R_h = H^{-1}\left(1+\Phi + {\cal H}^{-1}\left(\Phi'-\partial_\rho\Phi\right)\right) ~.
\eeq
In the long-wavelength limit $|k\eta|\ll 1$ the derivatives of $\Phi$ become subdominant and the radius of the horizon reads $R_h = H^{-1}(1+\Phi)$. As we mentioned earlier, in this limit $\Phi$ also becomes constant. Thus, including fluctuations we see that the horizon area is corrected as
\beq
\text{Area}(R_h) = 4\pi R_h^2 = \frac{4\pi}{H^2}\left(1+2\Phi\right) ~.
\eeq
Including perturbations, the Misner-Sharp mass is still expressed in terms of the horizon radius as
\beq
M(R_h) = \frac{R_h}{2G} ~,
\eeq
but now $R_h$ is given by \eqref{eq:ShiftedHor}. Again dropping derivatives of $\Phi$ the surface gravity is modified to be
\beq
\kappa = R_h^{-1}\left(1-\frac\varepsilon2\right) ~.
\eeq
Thus, we see that all the modifications in the horizon, mass and temperature can be captured by replacing the location of the horizon by $R_h$ as defined in \eqref{eq:ShiftedHor}.

Now we promote $M(R)$ to an operator whose leading contribution is proportional to the identity. Treating the fluctuations as quantized perturbations we identify
\beq\label{eq_identification}
K = \beta M(R_h) = \frac{\pi}{G_NH^2}\left(1+2\Phi\right) ~,
\eeq
where we expanded to leading order in small $\varepsilon$ and focused on long-wavelength perturbations. To leading order in $G_N$ this ansatz for the modular Hamiltonian assumes the density matrix of de Sitter space to be maximally mixed, as has been suggested in \cite{Dong:2018cuv,Lin:2022nss,Chandrasekaran:2022cip}. We note that this assumption is different from the one in \cite{Banks:2024cqo}, who argued in favor of a non-maximally mixed density matrix. We further assume that this identification \eqref{eq_identification} still holds when we include (quantized) perturbations. As we will see, this assumption correctly reproduces the relation for fluctuations in the modular Hamiltonian derived in \cite{Verlinde:2019ade} in the context of AdS/CFT.

Taking the vacuum expectation value we find that
\beq
\langle K \rangle = \frac{\pi}{G_NH^2} = S_{\rm dS} ~,
\eeq
because the one-point function of $\Phi$ vanishes.

We now follow \cite{Verlinde:2019ade} and capture fluctuations in the area of the causal diamond by defining the operator
\beq
\Delta K = K-\langle K\rangle\cdot {\bf 1} ~,
\eeq
which in our case is given by
\beq
\Delta K = 2S_{\rm dS}\Phi ~.
\eeq
This equation provides a direct relation between fluctuations of the modular Hamiltonian and the Newtonian potential $\Phi$ similar to the one that appeared in \cite{Verlinde:2019ade}. In fact, it exactly reproduces Eq. (67) of that paper.

We can now go one step further and use our results from Sec. \ref{sec:quantization} to compute the variance. We have
\beq \label{eq:FluctRela}
\langle\Delta K^2\rangle = 4S_{\rm dS}^2\langle\Phi ^2\rangle ~.
\eeq
Using \eqref{eq:RenormalizedVariance} we find this becomes equal to
\beq
\langle\Delta K^2\rangle = 2\varepsilon S_{\rm dS} \log(\mu/H) = 2\varepsilon\log(\mu/H)\langle K\rangle ~.
\eeq
We note that this result reproduces the proportionality of $\langle\Delta K^2\rangle \sim \langle K\rangle$ obtained by \cite{Verlinde:2019ade} in anti-de Sitter space and \cite{Banks:2024cqo} in de Sitter space for a CFT (see \cite{Perlmutter:2013gua,Nakaguchi:2016zqi,DeBoer:2018kvc} for earlier work), but with a prefactor different from one that depends on the renormalization group scale $\mu$. This is not surprising, as the theory we are studying is not conformally invariant. We should also note that the fluctuations of the modular Hamiltonian go to zero in the exact ($\varepsilon\to 0$) de Sitter limit, different from \cite{Banks:2024cqo}. As explained before, from a cosmologist's perspective this reflects the fact that the inflaton fluctuations become pure gauge in this limit. However, from the perspective of the modular Hamiltonian this is a consequence of our assumption that the density matrix of pure de Sitter is maximally mixed. We stress that \cite{Banks:2024cqo} (see also \cite{A:2025ajh}) made a different assumption for the density matrix.

So far we focused on causal diamonds that consisted of the entire static patch. We are also interested in smaller subregions for which the expectation value of the modular Hamiltonian is still given by the generalized entropy \cite{Jensen:2023yxy}. Using the same identification between the modular Hamiltonian and the Misner-Sharp mass we find that for a smaller subregion with a spherical entangling surface of radius $R$ to leading order in $\varepsilon$
\beq
K = \beta M(R) = \frac{\pi R^2}{G_N} = \frac{\pi a^2\rho^2}{G_N}\left(1-2\Phi\right) ~.
\eeq
We see that the fluctuations $\Delta K$ still obey the relation \eqref{eq:FluctRela} where the de Sitter entropy is replaced by $\text{Area}(R)/(4G_N)$. 
\beq
\langle\Delta K^2\rangle = \frac{\text{Area}(R)^2}{4G_N^2}\langle\Phi^2\rangle ~.
\eeq
Next, we study the implications of these fluctuations on observables.

\section{Impact of Quantum Fluctuations on Causal Diamonds} \label{sec:Diamonds}
In the previous section, we took an ansatz relating the modular Hamiltonian of a causal diamond in quasi-de Sitter space to the Misner-Sharp mass and reproduced the scaling of modular Hamiltonian fluctuations $\langle \Delta K^2\rangle \sim \langle K\rangle$, previously derived. Moreover, we were able to directly relate modular Hamiltonian fluctuations to a physical observable: primordial scalar fluctuations that lead to the temperature anisotropies in the CMB. This supports the idea that modular Hamiltonian fluctuations have observable effects.

We should be cautious however in directly comparing our results to earlier works on modular fluctuations. Crucially, the scalar fluctuations we study here are sourced by a non-zero matter sector. Instead, the modular Hamiltonian fluctuations studied for example in \cite{Verlinde:2019ade} were associated with (vacuum) fluctuations in the absence of matter. To make the relation between the two approaches clear we need to understand if the scalar fluctuations we study are also present in vacuum. 

One thing to note is that in the long-wavelength limit we studied only an s-wave perturbation survives. In fact, in Appendix \ref{app:Gauge} we explicitly relate the scalar perturbations $(\Phi,\Psi)$ to the $\ell=0$ modes in Regge-Wheeler gauge. Notably, the two physical degrees of freedom in the graviton are related to $\ell\geq 2$ modes and our s-wave perturbation can therefore not be associated to the radiative degrees of freedom of the gravitational field. Moreover, in the classical theory the $\ell = 0,1$ modes are absent in vacuum \cite{Zerilli:1970,Martel:2005ir}. Regardless, this does not prohibit such fluctuations to be generated off-shell in the quantum theory. It is therefore worthwhile to investigate the possibility of generating spherically symmetric perturbations in quantum gravity, either off-shell or through some different mechanism.

Having spelled out these subtleties, we will now demonstrate how the scalar fluctuations we study affect the behavior of null rays traveling along the horizon in a causal diamond. It is important to stress that the results in this section are completely independent of any assumptions about the modular Hamiltonian. 

Before we dive into the explicit computation, let us briefly explain the main idea. Consider a causal diamond in a background geometry defined in a suitable gauge-invariant manner. For instance, we can consider the worldline of a geodesic observer and specify a timelike interval defined by proper time $\tau=\tau_T-\tau_B$. Shooting null geodesics from $\tau_B$ to the future and from $\tau_T$ to the past, the intersection of the geodesics specifies the causal diamond. This definition is gauge invariant even in the presence of perturbations because we can always choose Fermi normal coordinates along the worldline of the observer, making sure the trajectory follows a geodesic.

We are now interested in the travel time of a photon that leaves at $\tau_B$, travels along the horizon, reflects off the bifurcation surface and returns to $\tau_T$. Fluctuations in the geometry will modify the size of the causal diamond which subsequently influences the travel time of photons. In particular, the fluctuations derived by \cite{Verlinde:2019xfb} scale as 
\beq \label{eq:VZ-scaling}
\text{Verlinde-Zurek Effect:} \quad \delta L^2 \sim \ell_p L  ~,
\eeq
This should be contrasted with the `conventional' expectation that $\delta L^2 \sim \ell_p^2$, which is likely unobservable. We will now compute the effect of scalar fluctuations to correlators of proper length and time on causal diamonds in quasi-de Sitter space. As we will see, for the spherically symmetric perturbations we study, the (proper) size of causal diamonds indeed scales according to \eqref{eq:VZ-scaling}.

\subsection{Proper Correlators in Cosmology}
Before we compute correlators of proper length and time, let us first consider a simpler measure of the size of a causal diamond. We will consider the proper distance between an observer at the pole of pure de Sitter space and the bifurcation surface.

In pure de Sitter space, the proper distance can be easily computed by making use of embedding coordinates. The de Sitter invariant distance $Z(x,y)$ between two points $(x,y)$ is defined as
\beq
Z(x,y) = H^2\eta_{AB}X^A(x)X^B(y) ~.
\eeq
Here $X^{A=0,\dots,4}$ are embedding coordinates with $\eta_{AB}$ the five-dimensional Minkowski metric (see Appendix \ref{app:Coordinates}). The proper distance is now given by \cite{Hartong:2004rra}
\beq
\zeta(x,y) = H^{-1} \arccos(Z(x,y)) ~.
\eeq
Working in static coordinates the proper distance between the pole at $(t,r)=(t,0)$ and a location $(t,r)$ is given by
\beq
\zeta = H^{-1}\arccos\left(\sqrt{1-H^2r^2}\right) ~.
\eeq
If we now expand around the location of the horizon, i.e. $r = H^{-1}(1-\delta r)$, we find
\beq
\zeta = H^{-1}\left( \frac\pi2 - \sqrt{2\delta r} + {\cal O}(\delta r)^{3/2} \right) ~.
\eeq
Crucially, interpreting $\delta r$ as a spherically symmetric perturbation we see that a shift \emph{linear} in the horizon leads to a correction in the proper distance that scales as a square root.

This simple observation has important consequences. In Sec. \ref{subsec:Thermo} we saw that in quasi-de Sitter space the radius of the horizon reads $R_h=H^{-1}(1+\Phi)$, which corresponds to $\rho_h=-\eta (1+2\Phi)$ in terms of the coordinate distance in planar coordinates.\footnote{Note that the relation between the radius of a sphere $R$ and the coordinate distance $\rho$ in perturbed quasi-de Sitter space is given by $R^2=a(\eta)^2\rho^2(1-2\Phi)$.} By using the coordinate transformation \eqref{eq:PlanarToStatic}, the correction to the horizon in static coordinates is given by $\delta r = 2\Phi$. Treating $\Phi$ as an operator and defining $\delta \zeta = \zeta - \frac{\pi}{2H}$ we see that the first non-zero correlation function is a four-point function\footnote{The one-point function $\langle\Phi\rangle$ vanishes.}
\beq \label{eq:ProperDistScaling}
\delta \zeta^2 = \sqrt{\langle \delta \zeta^4\rangle} = 4H^{-2}\sqrt{\langle \Phi^2 \rangle} \sim \ell_p/H~.
\eeq
Using \eqref{eq:RenormalizedVariance} we therefore see that fluctuations in the causal diamonds scale according to the Verlinde-Zurek effect. We will see that this scaling also holds for other causal diamonds in quasi-de Sitter space. Before we do so, we would like to mention that our motivation for studying these fluctuations comes from fluctuations in the modular Hamiltonian. However, our results involving fluctuations of the proper distance and time are independent from this perspective.

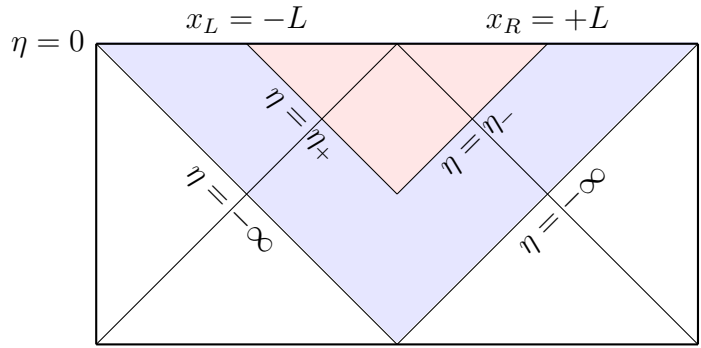
\begin{figure}[t]
\centering
\begin{tikzpicture}

\node[isosceles triangle,
    isosceles triangle apex angle=90,
    fill=blue!10,
    rotate=-90,
    minimum size =4cm] (T90)at (4,2.4){};

\node[isosceles triangle,
    isosceles triangle apex angle=90,
    fill=red!10,
    rotate=-90,
    minimum size =2cm] (T90)at (4,3.2){};

\draw[thick] (0,0) -- (8,0);
\draw[thick] (0,0) -- (0,4) node[pos=1,left]{$\eta  = 0$};
\draw[thick] (0,4) -- (8,4);
\draw[thick] (8,0) -- (8,4);

\draw (0,0) -- (4,4);
\draw (0,4) -- (4,0) node[pos=0.5,below, rotate=-45]{$\eta  = -\infty$};
\draw (4,0) -- (8,4) node[pos=0.5,below, rotate=+45]{$\eta  = -\infty$};;
\draw (4,4) -- (8,0);

\draw (2,4) -- (4,2) node[pos=0,above]{$x_L=-L$} node[pos=0.45,below,rotate=-45] {$\eta = \eta_+$};
\draw (4,2) -- (6,4) node[pos=1,above]{$x_R=+L$} node[pos=0.45,below,rotate=45] {$\eta = \eta_-$};

\end{tikzpicture}
\caption{Penrose diagram of de Sitter space, where we `unfolded' the spatial direction. The region shaded in blue is covered by the planar coordinates \eqref{eq:Planar}. The region shaded in red is covered by the boosted causal diamond coordinates \eqref{eq:MilneBoost}. The cutoff surface $\eta_0$ is located near $\eta=0$.}
\label{fig:CausalDiamond}
\end{figure}

We will now show how fluctuations influence the photon travel time. Instead of defining a diamond in the static patch, it is more appropriate to consider a causal diamond that is attached to future infinity. The reason is that, in cosmology, measurements about the inflationary phase of the universe are performed on a constant time slice. To this end, we consider a finite Milne-like wedge that is attached to a cutoff surface $\eta = \eta_0 \simeq 0$. This diamond can be defined in a gauge-invariant manner by considering the worldline of an observer located at $x=0$ that pierces the $\eta_0$ surface. Using Fermi normal coordinates we can easily make sure this trajectory is geodesic. At some particular coordinate time $\eta$ along this geodesic, we consider null geodesics that are emitted in opposite directions that also pierce the $\eta_0$ surface.\footnote{One can construct the $\eta_0$ surface in a gauge-invariant way by introducing a scalar function $\chi(x^\mu)$ such that $\chi(x^\mu)=\eta-\eta_0$ in the planar coordinates. The hypersurface is then defined by the equation $\chi(x^\mu)=0$.} For simplicity, we work along $y=z=0$. By fixing the proper time $\tau(\eta_0)-\tau(\eta)$ we can define the causal diamond in a gauge-invariant fashion, see Figure \ref{fig:CausalDiamond}.

The explicit coordinate transformation between this causal diamond and planar coordinates is given in \eqref{eq:MilneBoost}. As explained in Appendix \ref{app:Coordinates}, the interior of the causal diamond is described by the line element \eqref{eq:static-perturbed} (just as the entire Milne wedge), but the relation between planar and static coordinates is modified. As a result, the horizons of the causal diamond,\footnote{Because this finite causal diamond is obtained by boosting (an isometry) embedding coordinates (see Appendix \ref{app:Coordinates}), the boundary of the causal diamond is a Killing horizon on which the Killing vector becomes null, just as in the Milne wedge.}
located at $r_h = 1/H$, are given in planar coordinates by
\beq
\eta^2_\pm = L^2+ |\vec x|^2 \pm 2Lx ~,
\eeq
and, because we work along $y=z=0$, the bifurcation surface is given by $(\eta,x) = (-L,0)$. 

\begin{figure}[t]
\centering
\begin{tikzpicture}

\node[isosceles triangle,
    isosceles triangle apex angle=90,
    fill=red!10,
    rotate=-90,
    minimum size =2cm] (T90)at (3,-.8){};

\draw (0,0) -- (6,0) node[pos=1,right] {$~~~\eta=\eta_0$};

\draw (1,0)  -- (3,-2) node[pos=0,above]{$x_L=-(L+\eta_0)$};
\draw (3,-2)  -- (5,0) node[pos=1,above]{$x_R=+(L+\eta_0)$};
\draw[decorate,decoration={coil,aspect=0},red] (3,-2)  -- (1,0);
\draw[decorate,decoration={coil,aspect=0},red] (3,-2) -- (5,0);

\end{tikzpicture}
\caption{Finite causal diamond attached to a cutoff surface $\eta = \eta_0$. We consider the proper distance between a past directed null geodesic leaving the cutoff surface that reflects at the bifurcation surface and subsequently returns to the cutoff surface.}
\label{fig:CutoffDiamond}
\end{figure}
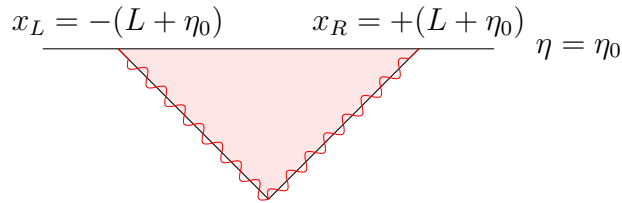

From the constant $\eta_0$ surface we now consider a photon that travels along the horizon, reflects at the bifurcation surface and then follows a future directed trajectory until it reaches $\eta_0$ again. The proper distance between these two events measured by an apparatus on the $\eta_0$ slice gives a measure of the size of a causal diamond, see Figure \ref{fig:CutoffDiamond}. As we will see, its fluctuations coincide with the scaling given earlier in \eqref{eq:ProperDistScaling}.

Before we compute this, we first consider the proper distance in the unperturbed geometry. The left and right spatial positions of the photons at the $\eta_0$ slice are given by
\beq
x_L = -(L + \eta_0) ~, \qquad x_R = +(L + \eta_0) ~.
\eeq
Without perturbations, the proper distance at $\eta=\eta_0$ is 
\beq
\zeta = \frac{1}{\eta_0H}\int_{x_L}^{x_R}\rmd x ~.
\eeq
The trajectory of the photon obeys $\rmd\eta^2 = \rmd x^2$ so this leads to
\beq
\zeta = \frac{2}{\eta_0H}\int^{\eta_0}_{-L}\rmd\eta = \frac{2}{\eta_0H}\left(L+\eta_0\right) ~.
\eeq
If we now include perturbations, there are three different effects that modify the proper distance we just computed. First, the expression of the proper distance is modified by terms of order ${\cal O}(\Phi)$ as the geometry is corrected. Second, the trajectory of the photon is modified by ${\cal O}(\Phi)$ terms in the metric. Third, the location of the bifurcation surface is corrected.

As we will now argue, for small perturbations the dominant effect is given by the shift in the location of the bifurcation surface as it scales as ${\cal O}(\sqrt{\Phi})$. We will therefore only focus on this effect. From the relations \eqref{eq:MilneBoost} we can derive the following relation between the Milne coordinates that cover the diamond and planar coordinates (along $y=z=0$)
\beq
H^2r^2-1 = \frac{\left(\eta^2-(L-x)^2\right)\left(\eta^2-(L+x)^2\right)}{4L^2\eta^2} ~.
\eeq
From this relation we see that the bifurcation surface at $r_h = H^{-1}(1+2\Phi)$ is shifted to $(\eta,x) = (\eta_h,0)$ with\footnote{It is important to note that we consider perturbations of the form \eqref{eq:static-perturbed} in Milne-like coordinates. Defining the bifurcation surface using the (conformal) Killing vector $\partial_t$ this leads to a change in the bifurcation surface of $r_h = H^{-1}(1+2\Phi)$.}
\beq
\eta_h = -L(1\pm2\sqrt{\Phi}) ~.
\eeq
As mentioned, the change in $\eta_h$ scales as ${\cal O}(\sqrt{\Phi})$ and therefore this effect dominates over the change in the expression of the proper distance and the trajectory of the null geodesic, which are only corrected at ${\cal O}(\Phi)$. The leading correction to the proper distance is thus given by
\beq\label{eq_LeadingProperDistance}
\zeta = \frac{2}{\eta_0H}\left(L+\eta_0\pm 2L\sqrt{\Phi}\right) ~.
\eeq
To remove the dependence on the cutoff surface $\eta_0$ we can now look at
\beq
\frac{\delta\zeta}{\zeta_0} = \pm2\sqrt{\Phi} ~,
\eeq
where $\zeta_0$ is defined as the proper distance without perturbations and $\delta \zeta$ the change to the proper distance. Treating $\Phi$ now as an operator we find that
\beq
\left(\frac{\delta\zeta}{\zeta_0}\right)^2 = \sqrt{\left\langle\left(\frac{\delta\zeta}{\zeta_0}\right)^4\right\rangle} = 4\sqrt{\langle\Phi^2\rangle} \sim \ell_pH~,
\eeq
in agreement with the VZ scaling, where we have used the scaling in the renormalized variance \eqref{eq:RenormalizedVariance}.

\begin{figure}[t]
\centering
\begin{tikzpicture}

\node[isosceles triangle,
    isosceles triangle apex angle=90,
    fill=red!10,
    rotate=-90,
    minimum size =2cm] (T90)at (3,-.8){};

\draw [gray,dashed] (3,-2) -- (3,2) node[pos=0.6,left] {$x=0$};

\draw (0,0) -- (6,0) node[pos=1,right] {$\eta=\eta_0$};
\draw (1,0)  -- (3,2) node[pos=1,above] {$\tau_T$};
\draw (3,2)  -- (5,0);
\draw (1,0)  -- (3,-2);
\draw (3,-2)  -- (5,0) node[pos=0,below] {$\tau_B$};
\draw[decorate,decoration={coil,aspect=0},red] (5,0)  -- (3,-2);
\draw[decorate,decoration={coil,aspect=0},red] (5,0) -- (3,2);

\end{tikzpicture}
\caption{Attaching a flat space region to the cutoff surface $\eta_0$ we consider the proper time $\tau = \tau_T-\tau_B$ measured by an observer at $x=0$ between a null geodesic leaving, reflecting at the edge of the diamond and returning.}
\label{fig:Hat}
\end{figure}
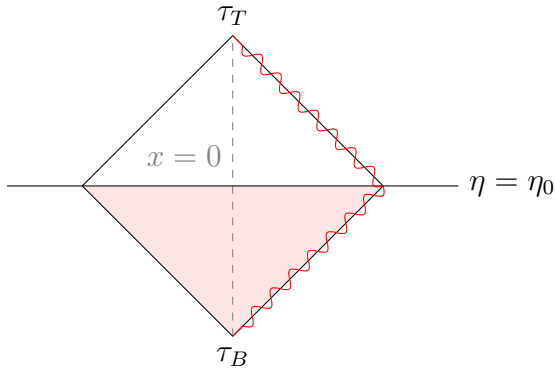

This proper distance can now be easily translated to a proper time by continuing the geometry to the future beyond the cutoff surface and attaching a flat space region. This can be thought off as a (crude) model describing the transition from inflation to a geometry without accelerated expansion and without perturbations. We then consider the same timelike observer following an $x=0$ trajectory that measures the proper time between the null geodesic leaving $x=0$ and returning at $x=0$ in the flat space geometry. To this end, it is useful to write the (unperturbed) planar metric in terms of the time coordinate $\tau=-H^{-1}\log(-H\eta)$ which is the proper time of an observer at $x=0$ (in the unperturbed geometry). In the flat space region, $\tau$ is the standard (proper) Minkowski time coordinate, see Figure \ref{fig:Hat}.
Because we are considering photons, the proper time $\tau = \tau_T-\tau_B$ is equal to the proper distance we computed before and we directly obtain
\beq
\tau = \frac{2}{\eta_0H}\left(L+\eta_0\pm2L\sqrt{\Phi}\right) ~.
\eeq
Clearly, treating $\Phi$ as an operator correlation functions of this proper time observable shows the same VZ scaling as the proper distance. As mentioned before, this result is independent of any assumptions about the modular Hamiltonian. Notably, our calculation implies that one way of understanding the VZ scaling is through quantizing an s-wave perturbation. To elaborate on this, we will discuss the VZ scaling in a more generic setup including other spacetimes in the next subsection.

\subsection{Comments on Other Spacetimes}
Before we end this section, let us comment on the effect of fluctuations in other spacetimes. In cosmology, we saw that the origin of the VZ scaling lies in the fact that a \emph{linear} shift in the horizon due to some spherically symmetric perturbation leads to a proper distance that is corrected by the square root of the perturbation. Is this still true in other spacetimes?

In anti-de Sitter space, the answer is yes. In fact, our computation of corrections to the proper distance in cosmology closely parallels the computation of the proper time of a photon traveling along the horizon of a causal diamond in anti-de Sitter space \cite{Verlinde:2019ade}.

More interesting is the situation in flat space. If we focus on the spherically symmetric perturbations we also studied in cosmology, we can easily see the same scaling holds true for flat space diamonds. Using the results of \cite{Casini:2011kv}, we can perform a coordinate transformation from the Minkowski metric to a metric conformal to the static or Milne patch of de Sitter space. Explicitly, let us write the flat space metric in polar coordinates as
\beq
\rmd s^2 = -\rmd \tau^2+\rmd \rho^2 + \rho^2\rmd\Omega_2^2 ~.
\eeq
Using the coordinate transformation
\beq
\bal \label{eq:FlatTrafo}
\tau &= \frac{L\sqrt{1-r^2/L^2}\sinh(t/L)}{1+\sqrt{1-r^2/L^2}\cosh(t/L)} ~, \\
\rho &= \frac{r}{1+\sqrt{1-r^2/L^2}\cosh(t/L)} ~,
\eal
\eeq
the flat space metric can be written as
\beq
\rmd s^2 = C(t,r)^2\left(f(r)\rmd t^2+f(r)^{-1}+r^2\rmd\Omega_2^2\right) ~,
\eeq
where
\beq
\bal
C(t,r)^{-1} &= 1+\sqrt{1-r^2/L^2}\cosh(t/L) ~,\\
f(r) &= 1-r^2/L^2 ~.
\eal
\eeq
We recognize this as the static metric up to a conformal factor $C(t,r)$. We thus find that this transformation maps the static patch of de Sitter space to a finite causal diamond with a spherical entangling surface of radius $r=L$ in flat space. We can now consider a linear perturbation to the bifurcation surface in de Sitter space and see how this maps to a change in the bifurcation surface in flat space.

Using \eqref{eq:FlatTrafo} we find that a spherically symmetric perturbation $r = L(1-\delta r)$ of the bifurcation surface at $t=0$ maps to $\tau=0$ and
\beq
\rho = L\left(1-\sqrt{2\delta r} + {\cal O}(\delta r)\right) ~.
\eeq
We see that the leading correction to the location of the bifurcation surface in flat space scales with the square root of the perturbation. Because we are working in flat space, this implies that the proper round-trip time of a photon leaving the observer at $\rho = 0$, traveling along the horizon and returning to the observer also has a correction that scales with the square root of the perturbation consistent with the VZ scaling.

As before, this assumes the presence of a gravitational s-wave perturbation that shifts the bifurcation surface of conformal-de Sitter space linearly. We'd like to stress that this assumption involves a choice of conformal Killing vector that specifies the perturbed causal diamond. It is possible, for example, to consider a conformal Killing vector for which the coordinate location of the bifurcation surface doesn't change in the perturbed metric.

Let us contrast the square root scaling of the round-trip time with the recent work \cite{Carney:2024wnp}. In that paper, it was demonstrated that quantized graviton fluctuations lead to uncertainty in the arm length of an interferometer that scales as $\delta L^2 \sim \ell_p^2$, i.e. the conventional way. From our analysis, we now understand the origin of this different scaling. Employing transverse traceless gauge, \cite{Carney:2024wnp} focused on quantizing the two physical polarizations of the graviton, i.e. the radiative degrees of freedom in the gravitational field. This leads to a correction to the proper distance of the arm that scales linearly with the perturbation. Instead, inspired by cosmology we quantized a perturbation that modifies the size of the horizon in a spherically symmetric manner. As we demonstrated in (quasi-)de Sitter space, anti-de Sitter space and flat space, this leads to a modification of the proper distance that scales with a square root. 

Using our cosmological setup with Hubble parameter $H=1/L$, let us now sketch in another way how quantizing these different degrees of freedom leads to a different scaling. As we saw, the radius and area of the cosmological causal diamond including perturbations (in the long-wavelength limit) are given by
\beq
R_h = L\left(1+\Phi\right) ~, \quad A(R_h) = 4\pi L^2(1+2\Phi) ~.
\eeq
By working to linear order in the perturbations we find that both the correction to the radius \emph{and} the correction to the area scale linearly with $\Phi$. If we are considering perturbations in transverse-traceless gauge, the effect of these perturbations is to modify the proper distance along a particular direction (transverse to the direction of propagation) while keeping the area fixed. Transverse-traceless perturbations therefore break isotropy by selecting a particular direction.\footnote{This is in line with the fact that (radiative) graviton fluctuations start at $\ell\geq 2$ in Regge-Wheeler gauge \cite{Maggiore:2018sht} and are not spherically symmetric. However, averaging over all angles, the net effect of the perturbations can still be spherically symmetric.} The change in the proper distance along this direction is given by $\delta L \sim L\Phi$. Viewing this as an operator, we quantize it to find
\beq
\text{Radius Quantization:} \quad \langle\delta L^2\rangle \sim L^2\langle\Phi ^2\rangle \sim \ell_p^2 ~,
\eeq
where we used the scaling in the variance \eqref{eq:RenormalizedVariance}. We refer to this scheme as `radius quantization'.

\begin{figure}[t]
\centering
\subfloat[]{{
\begin{tikzpicture} 

    \foreach \a in {0,5,...,355} {
        \fill[color=gray] ({2*cos(\a)},{2*sin(\a)}) circle (0.04);
    }
    \foreach \a in {0,5,...,355} {
        \fill[color=orange,rotate=45] ({1.5*cos(\a)},{(4/1.5)*sin(\a)}) circle (0.04);
    }
    
\end{tikzpicture}
}}
    \qquad
    \subfloat[]{{

\begin{tikzpicture}

    \foreach \a in {0,5,...,355} {
        \fill[color=gray] ({2*cos(\a)},{2*sin(\a)}) circle (0.04);
    }
    \foreach \a in {0,5,...,355} {
        \fill[color=orange] ({1.5*cos(\a)},{1.5*sin(\a)}) circle (0.04);
    }
    
\end{tikzpicture}
    }}
    \caption{The effect of a dimensionless perturbation $\Phi$ on a ring of point particles when; (a) the perturbation is transverse-traceless; (b) the perturbation is spherically symmetric. In (a), the change in the proper distance from the origin to the semi-minor/major axis scales linearly with the perturbation. In (b), the proper distance from the origin to the ring of particles scales with the square root of the perturbation.}
    \label{fig:Perturbations}
\end{figure}
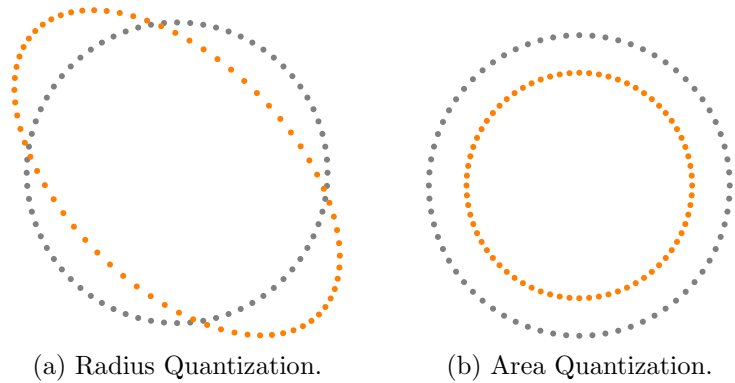

In contrast, when we focus on spherically symmetric perturbations we directly quantize the area of the diamond. It is therefore more natural to define the change in the area as $\Delta \text{Area} \sim L^2\Phi$ and quantize this perturbation. This leads to a change in the proper distance to the bifurcation surface of $\delta L = \sqrt{\Delta \text{Area}/(4\pi)}$, see Figure \ref{fig:Perturbations}, which leads to
\beq
\text{Area Quantization:} \quad \langle\delta L^2\rangle \sim L^2\sqrt{\langle\Phi ^2\rangle} \sim \ell_pL ~.
\eeq
Thus, we find that the VZ scaling is only obtained when we directly quantize the area whose dominant modification is given by spherically symmetric perturbations. Working in (dimensionless) Kruskal coordinates $x^\pm$ where the bifurcation surface is located at $x^+=x^-=0$, this scaling can be understood from the fact that the area operator is composite since $\delta r \sim \delta x^+\delta x^-$. In \cite{Verlinde:2019xfb} this leads to a correction to the proper time that agrees with the VZ scaling.

In inflationary cosmology, an s-wave mode is present due to the non-zero matter sector. However, as mentioned before, the presence of such a mode in the vacuum of quantum gravity is less clear. Additionally, while we have shown that correlators of proper distance and time exhibit the VZ scaling, we have not directly related these correlators to an interferometry setup. In particular, it would be worthwhile to better understand the relation between (a change in) the bifurcation surface and the (change in) the armlength of an interferometer. Clearly, this deserves future study.

\section{Discussion} \label{sec:Discussion}
In this work, we have studied quantum fluctuations of causal diamonds within cosmological backgrounds. Our main goal was to clarify the origin of the Verlinde-Zurek (VZ) scaling of modular Hamiltonian fluctuations. To this end, we considered scalar perturbations in a quasi-de Sitter spacetime characterized by a slow-roll parameter $\varepsilon$. 

Then, after reviewing the standard treatment of perturbations in inflationary cosmology, we proposed an ansatz for the modular Hamiltonian in terms of the Misner-Sharp mass (to leading order in $G_N$) which we promoted to an operator. This directly relates modular fluctuations to scalar perturbations. Using our ansatz, we computed the variance of the modular Hamiltonian inside the static patch: the largest causal diamond in de Sitter space. The variance scales as $\langle\Delta K^2\rangle \sim \varepsilon\langle K\rangle$, in agreement with the VZ effect \cite{Verlinde:2019ade}. An interesting result is that the fluctuations of the modular Hamiltonian vanish in the exact de Sitter limit $\varepsilon\to 0$. This is consistent with the expectation that pure de Sitter space has a maximally mixed density matrix, see however \cite{Banks:2024cqo}. By relating modular fluctuations to gauge-invariant scalar perturbations, which are directly linked to temperature anisotropies in the Cosmic Microwave Background, our work supports the idea that modular fluctuations are observable.

Independent of assumptions about modular Hamiltonians, we studied the effect of the spherically symmetric perturbation on proper length and proper time correlators in cosmology. We found that fluctuations in the proper length correlator scale as $\delta \zeta^2 \sim \ell_p/H$. This scaling agrees with \cite{Verlinde:2019xfb}.
Using a crude model describing the end of inflation in which cosmic expansion and perturbations terminate at a finite time, we demonstrated that proper time fluctuations obey the same scaling, again consistent with the VZ effect.

Interestingly, the geometric mean of the horizon radius and the Planck scale has also appeared as the scale relevant to certain quantum gravity effects in \cite{Marolf:2003bb,Freidel:2022ryr,Ong:2023lbr,Bousso:2023kdj,Banks:2024imv,Parikh:2024zmu}. The origin of this scaling seems to lie in a fundamental limitation on the number of degrees of freedom in a subregion given by the Bekenstein bound \cite{Bekenstein:1980jp} or its covariant version (the Bousso bound \cite{Bousso:1999xy, Bousso:2002ju}). In particular, in \cite{Freidel:2022ryr} the geometric mean of the Planck scale and inverse Hubble parameter arose due to a bound on the degrees of freedom in phase space, whereas in our work we directly studied fluctuations in the area (proportional to the entropy) of a cosmological causal diamond. To better understand the relation between the VZ effect and the setup of \cite{Freidel:2022ryr}, it would be interesting to examine the form of the density matrix used in \cite{Freidel:2022ryr}. This is a necessary step, because just satisfying a covariant entropy bound is insufficient to determine the precise form of the density matrix \cite{Banks:2024cqo}. Knowing the density matrix, it is then possible to compute the associated modular Hamiltonian and see if its fluctuations scale as $\langle \Delta K^2\rangle \sim \langle K\rangle$, as expected for the VZ effect.

Furthermore, using the fact that the coordinate system of a finite causal diamond in flat space is conformal to the static de Sitter metric, we provide evidence for the VZ scaling in flat space as well, assuming a perturbation that shifts the bifurcation surface of conformal-de Sitter space linearly. Our work therefore provides a bridge between inflationary cosmology and quantum gravity effects that are potentially observable in interferometric setups. Importantly, we emphasize that the scaling that we derived is independent of any assumptions about the modular Hamiltonian. Instead, it arises solely from imposing spherical symmetry on the quantized perturbations, in line with the notion that area is quantized in quantum gravity. In cosmology, this spherically symmetric perturbation is sourced by the scalar field driving inflation. However, it is less clear if such modes are present in the vacuum of a quantum gravity theory, for example in flat space.

Our work therefore suggests that s-wave metric fluctuations are at the heart of the VZ effect. A natural and interesting direction for future work is to study mechanisms to generate them. One promising approach is to make use of recent advances in asymptotic symmetries and soft modes. In particular, \cite{Bub:2024nan} found an asymptotic charge of a causal diamond in flat space that changes its size. A similar rich story of soft modes and asymptotic symmetries has been studied in cosmology, see e.g. \cite{Kehagias:2016zry}, and it would be worthwhile to see if there are more lessons about the VZ effect that can be extracted from cosmology.

Another future direction is to make the connection between our computation and a physical interferometer more clear. In particular, we studied the reflection of photons off a bifurcation surface instead of a physical mirror. Whether or not the bifurcation surface correctly models the mirror of an interferometer is an interesting question that we keep for future work. Regardless, we hope that our work has clarified the origin of the scaling of modular Hamiltonian fluctuations and proper time correlators and gives a fresh perspective on the important problem of finding observational signatures of quantum gravity.

\subsection*{Acknowledgments}
We are grateful to the members of the QuRIOS collaboration and especially Allic Sivaramakrishnan, Erik Verlinde and Kathryn Zurek for useful discussions and feedback. We also thank Patrick Draper for feedback and insightful remarks. LA and SEB are supported by the Heising-Simons Foundation under the “Observational Signatures of Quantum Gravity” collaboration grant 2021-2818. SEB is also supported by the U.S. Department of Energy under grant number DE-SC0019470. This work was performed in part at Aspen Center for Physics, which is supported by National Science Foundation grant PHY-2210452.

\appendix

\section{Coordinate Systems in de Sitter Space} \label{app:Coordinates}
In this Appendix we specify the different coordinate systems that we use. We will always be working on an FRW geometry close to de Sitter space and we therefore only consider coordinate transformations in that spacetime. Four-dimensional de Sitter space can be embedded into five-dimensional Minkowski space as:
\beq
\eta_{AB}X^AX^B = H^{-2} ~.
\eeq
Here, $A=(0,\dots,4)$, $\eta_{AB}$ is the Minowksi metric and $H$ is the Hubble parameter, which is constant in de Sitter space. From embedding coordinates, the line element is given by $\rmd s^2 = \eta_{AB}\rmd X^A\rmd X^B$.

\subsubsection*{Planar Coordinates}
Coordinates useful to describe inflation in which the expansion of de Sitter space is manifest are given by
\beq
\bal \label{eq:Planar}
X^0 &= \frac{H^{-2}-\eta^2+\vec x^2}{2\eta} ~, \\
X^4 &= \frac{H^{-2}+\eta^2-\vec x^2}{2\eta} ~, \\
X^i &= \frac{x^i}{H\eta} ~.
\eal
\eeq
Here $x^{i=1,2,3} = (x,y,z)$. The line element is
\beq
\rmd s^2 = \frac{1}{(H\eta)^2}\left(-\rmd \eta^2 + \rmd \vec x^2\right) ~.
\eeq
These coordinates cover half of the de Sitter Penrose diagram. $\eta\in(-\infty,0)$ and $x^i\in (-\infty,+\infty)$.

\subsubsection*{Static Coordinates}
In static coordinates, the time translation symmetry of de Sitter space is manifest. They are given by
\beq
\bal \label{eq:StaticCover}
X^0 &= \sqrt{H^{-2}-r^2}\sinh(t/\ell) ~, \\
X^4 &= \sqrt{H^{-2}-r^2}\cosh(t/\ell) ~, \\
X^i &= r\omega^i ~.
\eal
\eeq
Here $\omega^{i=1,2,3}=(\cos\theta,\cos\phi\sin\theta,\sin\theta\sin\phi)$ are coordinates on the unit two-sphere. The line element is
\beq
\bal
\rmd s^2 &= -f(r)\rmd t^2+f(r)^{-1}\rmd r^2 + r^2\rmd\Omega_2^2 ~, \\
f(r) &=1-H^2r^2 ~.
\eal
\eeq
Static coordinates cover a quarter of the Penrose diagram. $t\in (-\infty,+\infty)$ and $r\in (0,1/H)$.

We can also analytically continue these coordinates to cover the Milne wedge by sending $t\to t+\frac{i\pi}{2H}$. In that case, $r\in(1/H,\infty)$ is a timelike coordinate whereas $t$ becomes a spacelike coordinate that runs from $t\in(-\infty,+\infty)$.

\subsubsection*{Coordinate Relations}
We now derive the coordinate transformation between planar and static coordinates. The standard way of relating these coordinates is by identifying $X_{\rm static}^A = X^A_{\rm planar}$. We then find the relations
\beq
\bal
\sqrt{H^{-2}-r^2}\sinh(Ht) &= \frac{H^{-2}-\eta^2+|\vec x|^2}{2\eta} ~, \\
\sqrt{H^{-2}-r^2}\cosh(Ht) &= \frac{H^{-2}+\eta^2-|\vec x|^2}{2\eta} ~, \\
r\omega^i &= \frac{x^i}{H\eta} ~. 
\eal
\eeq
From here, we can derive
\beq\label{eq:PlanarToStatic}
H^2r^2 = \frac{|\vec x|^2}{\eta^2} ~.
\eeq
We see that this choice aligns the $x^i$ directions in planar coordinates with the planar coordinates on the two-sphere in static coordinates. This implies that the horizon in static coordinates at $r=1/H$ is given in planar coordinates by $\eta+|\vec x|=0$.

\subsubsection*{Causal Diamond Coordinates}
An alternative way of relating coordinates is by switching the identification of one of the embedding coordinates. This will result in a causal diamond that is attached to a finite region of future infinity. To this end, we choose to identify $(X^1,X^4)_{\rm static} = (X^4,X^1)_{\rm planar}$ and leave the rest of the identifications unchanged. In addition we send $t\to t+i\frac{\pi}{2H}$ to obtain Milne-like coordinates. This leads to the following relations
\beq
\bal \label{eq:MilneCoord}
\sqrt{r^2-H^{-2}}\cosh(Ht) &= \frac{H^{-2}-\eta^2+|\vec x|^2}{2\eta} ~, \\
\sqrt{r^2-H^{-2}}\sinh(Ht) &= \frac{x}{H\eta} ~, \\
r \omega^{i=(2,3)} &= \frac{x^{i=(2,3)}}{H\eta} ~, \\
r\omega^{i=1} &= \frac{H^{-2}+\eta^2-|\vec x|^2}{2\eta} ~.
\eal
\eeq
From here we find
\beq
r^2-H^{-2} = \frac{(H^{-2}-\eta^2+|\vec x|^2)^2-4x^2/H^2}{4\eta^2} ~.
\eeq
Thus, the horizon at $r_h=1/H$ is now located at
\beq
\eta^2_\pm = H^{-2}+|\vec x|^2 \pm 2x/H ~.
\eeq
Finally, we can also consider causal diamonds attached to future infinity with an extend given by $x\in (x_L,x_R) = (-L,L)$ by boosting in embedding space. This analogous relations in anti-de Sitter space can be found in \cite{Casini:2011kv}. We perform a boost in embedding coordinates as
\beq
\bal
X^0 &\to (X^0)' = \cosh\alpha\, X^0 - \sinh\alpha\, X^4 ~, \\
X^4 &\to (X^4)' = \cosh\alpha\, X^4 - \sinh\alpha\, X^0 ~.
\eal
\eeq
Because this is an isometry, the line element remains unchanged. Identifying $e^{-\alpha} = HL$ we see that the relation between Milne coordinates and planar coordinates is given by
\beq
\bal \label{eq:MilneBoost}
\sqrt{H^2r^2-1}\cosh(Ht) &= \frac{L^2-\eta^2+|\vec x|^2}{2\eta L} ~, \\
\sqrt{H^2r^2-1}\sinh(Ht) &= \frac{x}{\eta} ~, \\
2Hr \cos\theta &= \frac{L^2+\eta^2-|\vec x|^2}{\eta L} ~, \\
Hr \cos\phi\sin\theta &= \frac{y}{\eta} ~, \\
Hr \sin\phi\sin\theta &= \frac{z}{\eta} ~.
\eal
\eeq
The horizon $r_h=1/H$ expressed in planar coordinates is
\beq
\eta_\pm^2 = L^2 +|\vec x|^2 \pm 2Lx ~,
\eeq
with the bifurcation surface located at $(\eta,x) = (-L,0)$. The regions covered by these coordinates are indicated in Figure \ref{fig:CausalDiamond}.

\section{Relation to Regge-Wheeler Gauge} \label{app:Gauge}
In this Appendix, we show how the choice of conformal Newtonian gauge made in \eqref{eq:PlanarPerturbed} relates to the spherically symmetric perturbations in Regge-Wheeler gauge. We start with a static spherically symmetric background metric.
\beq \label{eq:SpherSymMet}
\rmd s^2 = -f(r)\rmd t^2 +f(r)^{-1}\rmd r^2 + r^2\rmd\Omega_2^2 ~.
\eeq
We now consider a general perturbation $g_{\mu\nu} = \bar g_{\mu\nu} + h_{\mu\nu}$, where $\bar g_{\mu\nu}$ is given by \eqref{eq:SpherSymMet}. Under a diffeomorphism $x^\mu \to x^\mu + \xi^\mu$ the metric perturbation transforms as
\beq \label{eq:diffeo}
h_{\mu\nu} \to h_{\mu\nu} - (\bar\nabla_\mu\xi_\nu+\bar\nabla_\nu\xi_\mu) ~.
\eeq
where $\bar\nabla_\mu$ is the covariant derivative with respect to the background geometry $\bar{g}_{\mu\nu}$. We are only interested in spherically symmetric perturbations. For this purpose, it is sufficient to focus on diffeomorphisms that are only non-zero in the $(t,r)$ directions: $\xi=\xi^t\partial_t + \xi^r\partial_r$ and $\xi^\mu = \xi^\mu(t,r)$. In a decomposition in spherical harmonics this selects just the $\ell=0$ modes. A complete decomposition can be found in \cite{Maggiore:2018sht}.

Following standard conventions, we write the metric perturbations as
\beq
h_{\mu\nu} = 
\begin{pmatrix}
h_{tt}(t,r) & h_{Rt}(t,r) & \, & \, \\
h_{Rt}(t,r) & h_{L0}(t,r) & \, & \, \\
\, & \, & h_{T0}(t,r) & \, \\
\, & \, & &\, & \, h_{T0}(t,r)\sin^2\theta  
\end{pmatrix} ~.
\eeq
Under the diffeomorphism \eqref{eq:diffeo} the metric perturbations transform as
\beq
\bal
h_{tt} &\to h_{tt} +\xi_rf(r)\partial_rf(r) -2\partial_t\xi_t ~, \\
h_{Rt} &\to h_{Rt} + \frac{\partial_rf(r)}{f(r)}\xi_t - \partial_r\xi_t - \partial_t\xi_r ~, \\
h_{L0} &\to h_{L0} - \frac{\partial_rf(r)}{f(r)}\xi_r - 2\partial_r\xi_r ~, \\
h_{T0} &\to h_{T0} - 2rf(r)\xi_r ~.
\eal
\eeq
From these transformations we can deduce the two (physical) scalar perturbations that are invariant. It can be checked that the two invariants are given by
\beq
\bal \label{eq:invariants}
\chi &= h_{L0} + \frac{2f(r)+r\partial_rf_r(r)}{2r^2f(r)^2}h_{T0} - \frac{\partial_rh_{T0}}{rf(r)} ~, \\
\omega &= h_{L0} - \frac{2}{f(r)^2}\left(h_{tt}-\frac{f(r)}{\partial_rf(r)}\partial_rh_{tt}\right) - \frac{(\partial_rf(r))^2-2f(r)\partial_r^2f(r)}{2rf(r)^2\partial_rf(r)}h_{T0} \\
&+ \frac{2}{rf(r)^2\partial_rf(r)}\partial_t^2h_{T0} - \frac{4}{f(r)\partial_rf(r)}\partial_th_{Rt} ~.
\eal
\eeq
We can now use the freedom to choose the diffeomorphism $\xi_\mu$ in a way we like. A particularly simple choice is to pick $\xi_t$ and $\xi_r$ such that $h_{T0}=h_{Rt}=0$, which corresponds to (generalized) Regge-Wheeler gauge \cite{Kallosh:2021ors}. The remaining spherically symmetric perturbations $h_{tt}$ and $h_{L0}$ do not correspond to radiative degrees of freedom. Instead, they have the effect of shifting the horizon \cite{Zerilli:1970,Martel:2005ir}.

Now specifying the background to be de Sitter space, we want to obtain conformal Newtonian gauge in static coordinates as given in \eqref{eq:static-perturbed}. First of all, we will choose $\xi_t$ and $\xi_r$ to shift 
\beq
\bal
h_{T0} &\to -2r^2\Phi~, \\
h_{Rt} &\to \frac{2Hr(\Phi+\Psi)}{1-H^2r^2} ~.
\eal
\eeq
There are the values these metric perturbations take in Newtonian gauge. After this gauge transformation, the remaining two metric perturbations $h_{tt}$ and $h_{L0}$ are now a function of $h_{tt}(t,r)$, $h_{L0}(t,r)$, $\Psi(t,r)$ and $\Phi(t,r)$. Since $h_{tt}(t,r)$ and $h_{L0}(t,r)$ were arbitrary perturbations we can redefine them to yield
\beq
\bal
h_{tt} &\to -2(H^2r^2\Phi+\Psi) ~, \\
h_{L0} &\to -\frac{2(H^2r^2\Psi+\Phi)}{(1-H^2r^2)^2} ~.
\eal
\eeq
We now have reduced the four degrees of freedom in the metric to the two physical ones. This yields the metric \eqref{eq:static-perturbed}. Plugging these components into \eqref{eq:invariants} we find the direct relation between the invariants ($\chi,\omega$) and $(\Phi,\Psi)$.

\bibliographystyle{utphys}

\end{document}